\documentclass[a4paper,11pt]{article}

 \usepackage{amsfonts,amsthm,amssymb}
 \usepackage{latexsym, color,epsf}

\newcommand{\eq}{\begin{equation}}
\newcommand{\en}{\end{equation}}
\newcommand{\eqa}{\begin{eqnarray}}
\newcommand{\ena}{\end{eqnarray}}

\begin{document}

  \setlength{\unitlength}{1mm}

  \thispagestyle{empty}

 \begin{center}
  {
  \bf  Virtual Extension of Temperley--Lieb Algebra
   }

  \vspace{.3cm}

  Yong Zhang ${}^a$\footnote{yzhang@nankai.edu.cn},
  Louis H. Kauffman ${}^b$\footnote{kauffman@uic.edu}
   and Mo-Lin Ge ${}^a$\footnote{geml@nankai.edu.cn}\\[.2cm]

 ${}^a$ Theoretical Physics Division, Chern Institute of Mathematics\\
  Nankai University, Tianjin 300071, P. R. China\\[.2cm]

 ${}^b$ Department of Mathematics, Statistics and Computer Science\\
  University of Illinois at Chicago, 851 South Morgan Street\\
  Chicago, IL, 60607-7045, USA \\[0.1cm]

\end{center}

\vspace{0.1cm}

\begin{center}
\parbox{13.5cm}{
\centerline{\small  \bf Abstract}  \noindent

 The virtual knot theory is a new interesting subject in the recent
 study of low dimensional topology. In this paper, we explore the
 algebraic structure underlying the virtual braid group and call it
 the virtual Temperley--Lieb algebra which is an extension of the
 Temperley--Lieb algebra by adding the group algebra of the
 symmetrical group. We make a connection clear between the Brauer
 algebra and virtual Temperley--Lieb algebra, and show the
 algebra generated by permutation and its partial transpose to be
 an example for the virtual Temperley--Lieb algebra and its important
 quotients.

 }

\end{center}

\begin{tabbing}


\end{tabbing}

\newpage

\section{Introduction}

  We recall the historical development of knot theory \cite{kauffman0}
  since Jones's seminal work \cite{jones1, jones2, jones3}.
  A braid representation in terms of the Temperley--Lieb (TL) algebra
  \cite{lieb, kauffman1} leads to the formulation of the Jones polynomial,
  while a braid representation using the two-parameter Hecke algebra
  derives  the HOMFLY polynomial \cite{homfly}. The Birman--Wenzl algebra
  \cite{wenzl1, murakami} is a generalization of the skein relations
  of the Kauffman two-variable polynomial  \cite{kauffman2} and it maps
  to the Brauer algebra \cite{brauer, wenzl2} in analogy to the map of
  the Hecke algebra to the group algebra of the symmetric group.

  In the recent study of low dimensional topology, the virtual knot
  theory becomes an interesting topic and this adds virtual crossings to
  knot theory which are a representation of the symmetric group, see
  \cite{kauffman3, kauffman4, kauffman5, kauffman6, dye1,
  manturov}.  As a closure of classical crossings is a knot (link),
  a closure of classical and virtual crossings leads to a virtual knot (link).
  The virtual braid group is generated by classical and virtual
  crossings.  In the present paper, we explore the algebraic structure
  underlying the virtual braid group and name it the virtual TL
  algebra which is a virtual generalization of the TL
  algebra by involving virtual crossings.

  In the literature, the virtual TL algebra has been implicitly
  argued \cite{dye2, kauffman7} and independently proposed
  \cite{yong}. As an extension of  \cite{yong}, in this paper we will
  study the virtual TL algebra from a pure algebraic viewpoint,
  i,e., the mixed relations between TL idempotents and virtual crossings
  are determined by a presumed requirement that virtual braids can be
  represented in the virtual TL algebra. Hence we will be able to make the
  connection clear between the Brauer algebra and virtual TL algebra.
  The Brauer algebra \cite{brauer, wenzl2} is generated by usual TL
  idempotents plus an operator that behaves like a permutation,
  denoting an algebra of all possible connections between $n$ points
  and $n$ points in the graphical sense.

  The plan of this paper is organized as follows. Section 2 derives those
  algebraic relations for defining the virtual Temperley--Lieb algebra and
  its important quotients. Section 3 builds the connection between the
  virtual TL algebra and the Brauer algebra, and shows that the algebra
  generated by permutation and its partial transpose \cite{yong} is an example
  for the virtual TL algebra and its important quotients. Last section remarks
  applications of the virtual TL algebra and virtual braid group to quantum
  information \cite{yong1, yong2}.

 \section{The virtual TL algebra and its quotients}

  After the sketch of the virtual braid group and its quotients,  we define
  the virtual TL algebra and its two important quotients:
  the welded virtual TL algebra and unrestricted virtual TL algebra.

 \subsection{The virtual braid group and its quotients}

 The virtual braid group $VB_n$ \cite{kauffman3, kauffman4, kauffman5,
kamada} is an extension of the classical braid group $B_n$ by
involving virtual crossings. Classical crossings $\sigma_i$,
generators of  $B_{n}$, satisfy the braid group relation called
$``BGR"$, \eqa \label{bgr}
  BGR:\,\, \sigma_{i}  \sigma_{i+1} \sigma_{i} &=& \sigma_{i+1} \sigma_{i} \sigma_{i+1},
  \qquad i=1, \cdots, n-1, \nonumber\\
   \sigma_{i} \sigma_{j} &=& \sigma_{j} \sigma_{i}, \qquad  j \neq i \pm
   1.
 \ena
Virtual crossings $v_i$ form a representation of the group algebra
of the symmetric group and satisfy the virtual crossing relation
called $``VCR"$, \eqa \label{vbgr1}
 VCR:\,\, v_i^2 &=& 1\!\! 1, \qquad  v_i v_{i+1} v_i = v_{i+1} v_i v_{i+1},
 \nonumber\\
 v_i v_j &=& v_j v_i, \qquad j \neq i \pm 1,
\ena  the symbol $1\!\! 1$ denoting the identity operator. Besides
$``BGR"$ and $``VCR"$, virtual crossings $v_i$ and classical
crossings $\sigma_j$ have to satisfy the mixed relations called
``$VBR$", \eqa \label{vbgr2}
 VBR: \,\,  \sigma_i v_j &=& v_j \sigma_i, \qquad   j \neq i \pm 1,
 \nonumber\\ v_i \sigma_{i+1} v_i &=& v_{i+1} \sigma_i v_{i+1}.
\ena

The move with two classical crossings and one virtual crossing is a
forbidden move in the virtual knot theory
 \cite{kamada,FRR, KANENOBU, NELSON}. There are two types of forbidden moves:
 the first one denoted by $(F_1)$ and the second denoted by $(F_2)$, \eq
\label{forbidden} (F_1): v_{i} \sigma_{i+1} \sigma_{i} =
\sigma_{i+1} \sigma_i v_{i+1}, \qquad (F_2): \sigma_i \sigma_{i+1}
v_{i} = v_{i+1} \sigma_{i} \sigma_{i+1}. \en  The first forbidden
move $(F_1)$ preserves the combinatorial fundamental group, as is
not true for the second forbidden move $(F_2)$. This makes it
possible to take an important quotient of the virtual braid group
$VB_n$. The welded braid group $WB_n$ \cite{kamada} satisfies the
same isotopy relations as the $VB_n$ group but allows the forbidden
move $(F_1)$. The unrestricted virtual braid group $UB_n$ allows
both forbidden moves $(F_1)$ and $(F_2)$ although any classical knot
can be unknotted in the virtual category if we allow both forbidden
moves \cite{FRR,KANENOBU, NELSON}.

\subsection{The virtual Temperley--Lieb algebra $vTL_n$ }

The virtual TL algebra $vTL_n$ is a virtual extension of the TL
algebra $TL_n$ by adding virtual crossings $v_i$ (\ref{vbgr1}). The
Temperley--Lieb relation called $``TLR(\lambda)"$, which are
satisfied by generators $E_i$ of the TL algebra $TL_n(\lambda)$ with
the loop parameter $\lambda$, are given by \eqa \label{tl}
TLR(\lambda):\,\,
 E_i^2 &=& \lambda E_i, \qquad (E_i)^\dag=E_i,\,\,\, i=1,\ldots,n-1, \nonumber\\
 E_i E_{i\pm1} E_i &=& E_i, \qquad E_i E_j=E_j E_i, \,\,\, |i-j|>1.
  \ena
Besides $VCR$ and $TLR(\lambda)$, $E_i$ and $v_i$ have to satisfy
the mixed relations which make it possible to represent virtual
braids in the virtual Temperley--Lieb algebra.

A linear representation $\rho$  of the $VB_n$ group assumes a form
in terms of $E_i$ and $v_i$, \eq \label{rho} \rho: vTL_n \to VB_n,
\qquad \rho_i=a +b E_i + c v_i, \en where $a, b, c$ are parameters
to be determined and this representation has to satisfy the $BGR$
and $VBR$ relations. The $VBR$ relation is a linear formulation of
braids $\sigma_i$, and this leads to a mixed relation between $E_i$
and $v_i$ called $``VEV"$ which is irrelevant with the parameters
$a, b, c$, \eqa \label{vev}
 VEV: \,\,  v_i E_{i+1} v_i &=& v_{i+1} E_i v_{i+1},\,\,\,
 i, j=1,2,\cdots n-1, \nonumber\\ E_i v_j &=& v_j E_i,
 \qquad  j\neq i \pm 1.
  \ena

Note that the $VEV$ relation together with the $VCR$ relation
derives $E_{i+1}$ in terms of $E_i$, $v_i, v_{i+1}$ by $E_{i+1}=v_i
v_{i+1} E_i v_{i+1} v_i$. This means that the $vTL_n$ algebra can be
generated by the idempotent $E_1$ and a set of virtual crossings
$v_i$. As a kind of heritage, the virtual braid group $VB_n$ can be
generated by the crossing $\sigma_1$ and the set of virtual
crossings $v_i$, see \cite{kauffman8}.

With the help of the $VEV$, $TLR(\lambda)$ and $VCR$ relations, the
$BGR$ relation in terms of the linear representation $\rho$
(\ref{rho}) has a simplified form called the $(vTL)$ relation, \eqa
\label{vtl} (vTL): && 0= (a^2 b+ a b^2 \lambda +b^3)(E_i -E_{i+1})
+a^2 c (v_i -v_{i+1})
\nonumber\\
 && + abc (E_i v_i +v_i E_i -E_{i+1} v_{i+1} -v_{i+1} E_{i+1})
 + b^2 c \sum_{j=0}^2 [F]_j
\ena where the symbols $[F]_0$, $[F]_1$ and $[F]_2$ are given by\eqa
[F]_0 &=& E_i v_{i+1} E_i -E_{i+1} v_i E_{i+1},
 \nonumber\\
  \lbrack F \rbrack_1 &=& v_i E_{i+1} E_i -E_{i+1} E_i v_{i+1}, \nonumber\\
  \lbrack F \rbrack_2 &=& E_i E_{i+1} v_i - E_{i+1} E_i v_{i+1}.
\ena

The $(vTL)$ relation can be further simplified under specific
circumstances. As $b=0, ac \neq 0$, it leads to $v_i=v_{i+1}$ which
is forbidden. For convenience, $b$ is non-vanishing in the
following. As $c=0, ab\neq 0$, it is reduced to an equation of $a,
b$, solved by setting $a=1$, \eq \label{ab} a^2 b + ab^2 \lambda
+b^3=0, \qquad a=1, \,\, b_\pm=-\frac 1 2 (\lambda\mp \sqrt{
\lambda^2-4}\,\,), \en  which shows that $\rho_\pm=1\!\! 1 +b_\pm E$
is a braid representation. As $a=0, bc\neq 0$, it has a simplified
form, \eq b (E_i-E_{i+1}) +c \sum_{j=0}^2 [F]_j=0. \en

Hence {\em the virtual TL algebra $vTL_n$ in the present paper is an
algebra generated by TL idempotents $E_i$ satisfying $TLR(\lambda)$
and virtual crossings $v_i$ satisfying $VCR$, and $E_i, v_i$ have to
satisfy the mixed relations: $VEV$ and $(vTL)$.}

 \subsection{The welded and unrestricted virtual TL algebras}

The welded and unrestricted virtual TL algebras $wTL_n$ and $uTL_n$
are two important quotients of the virtual TL algebra $vTL_n$ in
view of the fact that the virtual braid group $VB_n$ has two
important quotients including the welded and unrestricted virtual
braid groups $WB_n$ and $UB_n$.

The first forbidden move $(F_1)$, see (\ref{forbidden}), in terms of
the braid $\rho$ (\ref{rho}) has the form called the $(FF_1)$
relation,
 \eqa
(FF_1):  a^2 (v_i -v_{i+1})=- a b (v_i E_i -E_{i+1} v_{i+1}+ v_i
E_{i+1}- E_i v_{i+1}) - b^2 [F_1]
 \ena
 which is independent of the parameter $c$. With the help of this
 $(FF_1)$ relation, the $(vTL)$ relation is replaced by the following
 $(wTL_1)$ relation,
 \eqa
(wTL_1): & & (a^2 b + a b^2 \lambda +b^3) (E_i -E_{i+1})+ b^2 c
([F_0] + [F_2])\nonumber\\
 & & + abc (E_i v_i -v_{i+1} E_{i+1} -v_i E_{i+1} + E_i v_{i+1})
 =0. \ena
As $c=0, ab\neq 0$, the $(wTL_1)$ relation derives the same $a, b$
as (\ref{ab}). As $a=0, bc\neq 0$, the $(wTL_1)$ and $(FF_1)$
relations have the simplified forms, \eq b (E_i-E_{i+1}) + c
([F]_0+[F]_2)=0, \qquad [F]_1=0.
 \en

Therefore, {\em the welded TL algebra $wTL_n$ is generated by TL
idempotents $E_i$ satisfying $TLR(\lambda)$ and virtual crossings
$v_i$ satisfying $VCR$, and $E_i, v_i$ have to satisfy the mixed
relations $VEV$, $(FF_1)$ and $(wTL_1)$.}

Now we derive the $(FF_2)$ relation from the second forbidden move
$(F_2)$, see (\ref{forbidden}), using the braid $\rho$ (\ref{rho}),
\eq (FF_2): a^2 (v_i-v_{i+1})=-a b (E_i v_i -v_{i+1} E_{i+1}
+E_{i+1} v_i-v_{i+1} E_i) -b^2 [F_2] \en which is substituted into
the $(vTL)$ relation to derive the $(wTL_2)$ relation, \eqa (wTL_2):
&& (a^2 b+ a b^2 \lambda
+b^3 ) (E_i -E_{i+1}) + b^2 c ([F]_0 + [F]_1) \nonumber\\
 & & + abc (v_i E_i -E_{i+1} v_{i+1} -E_{i+1} v_i +v_{i+1} E_i) =0.
\ena As $c=0, ab\neq 0$, the same $a, b$ as (\ref{ab}) are required.
As $a=0, bc\neq 0$, the $(FF_2)$ and $(wTL_2)$ relations have the
following forms, \eq b (E_i -E_{i+1}) + c ([F]_0+[F]_1)=0, \qquad
[F]_2=0. \en

As a result, {\em the unrestricted TL algebra $uTL_n$ is generated
by TL idempotents $E_i$ satisfying $TLR(\lambda)$ and $v_i$
satisfying $VCR$, and $E_i, v_i$ have to satisfy the mixed relations
$VEV$, $(FF_l)$ and $(wTL_l)$, $l=1,2$.} As $c=0, ab\neq 0$, the
parameters $a, b$ are fixed by (\ref{ab}). As $a=0, bc\neq 0$, the
unrestricted TL algebra $uTL_n$ satisfies the simplified mixed
relations,
 \eq [F]_1=[F]_2=0, \qquad b (E_i -E_{i+1}) + c [F]_0 =0. \en

  \section{The Brauer algebra and virtual TL algebra}

We study the relationship between the virtual TL algebra and Brauer
algebra, and present an example for the virtual TL algebra and its
quotients which is an algebra generated by permutation and its
partial transpose.

 \subsection{The Brauer algebra as an virtual extension of the TL algebra}

 The Brauer algebra $D_n(\lambda)$ \cite{brauer} with the loop
 parameter $\lambda$ is generated by $TL$ idempotents $E_i$ and virtual
 crossings $v_i$, $i=1,\cdots, n-1$, which satisfy the  mixed relations
 given by \eqa \label{brauer}
 & & (ev/ve): E_i v_i = v_i E_i =E_i, \qquad
  E_i v_j = v_j E_i, \qquad j\neq i \pm 1,
 \nonumber\\
 & &(vee): v_{i\pm1} E_{i} E_{i\pm1} = v_{i} E_{i\pm1},
 \qquad (eev): E_{i\pm1} E_i v_{i\pm1}=E_{i\pm 1} v_i.
 \ena
 Obviously, the $TLR(\lambda)$, $VCR$ and $VEV$ relations for
 defining the virtual TL algebra are automatically satisfied in
 the Brauer algebra, and so we only need to
 examine under which conditions the $(vTL)$, $(wTL_l)$ and $(FF_l)$,
 $l=1,2$ relations are satisfied in the Brauer algerba.

 The three symbols $F_0$, $F_1$ and $F_2$ in the $(vTL)$ relation have
 the explicit forms in the Brauer algebra, \eq [F]_0=E_i-E_{i+1}, \,\,
  [F]_1=v_{i+1} E_i -E_{i+1} v_i,\,\, [F]_2=E_i v_{i+1}-v_i E_{i+1}, \en
 They are forbidden to be vanishing because $[F]_j=0$ leads to $E_{i}=E_{i+1}$
 which is not allowed in the Brauer algebra. The $(vTL)$ relation
 is replaced by an equation in the Brauer algebra,
 \eq \label{brvtl} a^2 c (v_i -v_{i+1}) + b ( a^2 + a b \lambda+b^2
 +c (2 a + b )) [F]_0 + b^2c ([F_1]+[F_2])=0, \en
 which has a solution given by (\ref{ab}) as the objects including
 $v_i-v_{i+1}$, $[F]_1$, $[F]_2$ and $[F]_3$ are regarded as linearly
 independent. Hence {\em the Brauer algebra $D_n(\lambda)$ is an example
 for the virtual TL algebra $vTL_n(\lambda)$ as (\ref{brvtl}) can be
 satisfied}.

 Similarly, the $(wTL_1)$ relation for defining the welded virtual TL algebra
 has the form in the Brauer algebra, \eq b (a^2
+a b \lambda +b^2 + c (a+b)) [F]_0 + b c (a+b) [F]_2=0, \en which
determines $a=-b$ and $\lambda=2$ as $[F]_0$ and $[F]_2$ are
linearly independent. The $(FF_1)$ relation has a simplified form
denoted by $<F_2>$ similar to the second forbidden move
(\ref{forbidden}), \eq <F2>:\,\,\, E_i^\ast E_{i+1}^\ast v_i
=v_{i+1} E_i^\ast E^\ast_{i+1}, \qquad E_i^\ast=1\!\! 1-E_i, \en
where $E_i^\ast$ is permutation-like $E_i^2=1\!\! 1$ at $\lambda=2$
and forms a braid representation since (\ref{ab}). Hence {\em the
Brauer algebra $D_n(2)$ modulo the forbidden move $<F_2>$ is the
welded vitual TL algebra $wTL_n(2)$ with $a=b=-1$}.

Furthermore, the $(wTL_2)$ relation has a simplified form in the
Brauer algebra \eq b (a^2 +a b \lambda +b^2 + c (a+b)) [F]_0 + b c
(a+b) [F]_2=0 \en which specifies $a=-b$, $\lambda=2$ and rewrites
the $(FF_2)$ relation into the form  denoted by $<F_1>$ similar to
the first forbidden move (\ref{forbidden}): \eq <F1>:\,\,\,\, v_i
E^\ast_{i+1} E^\ast_i=E^\ast_{i+1} E^\ast_i v_{i+1}. \en Therefore,
{\em the Brauer algebra $D_n(2)$ modulo both forbidden moves $<F_1>$
and $<F_2>$ is the unrestricted virtual TL algebra $uTL_n(2)$ with
$a=b=-1$.}

 \subsection{Permutation and its partial transpose}

 ${\cal H}_1$ and ${\cal H}_2$ are two independent $d$-dimensional Hilbert
 spaces with bases $\{|i\rangle\}$ and $\{|j\rangle\}$, and the tensor product
 $|i\rangle \otimes | j \rangle$ denoted by $|ij\rangle$ are product bases
 of ${\cal H}_1 \otimes {\cal H}_2$. The partial
transpose operator $\Theta_2$ acts on the operator product $A\otimes
B$ and only transforms indices belonging to the bases of the second
Hilbert space ${\cal H}_2$, namely $\Theta_2(A\otimes B)=A\otimes
B^T$. When the bases of ${\cal H}_2$ are fixed, the symbol $B^T$
denotes the matrix transpose.

The permutation operator $P$ has the form by $P=\sum_{i,j=1}^d |i j
\rangle \langle j i |$ which satisfies $P |\xi \eta\rangle=|\eta\xi
\rangle$.  With the partial transpose $\Theta_2$ acting on the
permutation $P$, we introduce a new operator $P_\ast$ given by \eq
\label{past} P_\ast=\Theta_2\circ P=\sum^d_{i, j=1} (|i
\rangle\otimes \langle j|) (|j\rangle \otimes \langle i |)^T, \qquad
P_\ast |\xi \eta\rangle=
 \sum_{i=1}^d |ii\rangle \delta_{\xi \eta}.
\en The algebra generated by the permutation $P$ and its partial
transpose $P_\ast$ is found to be the Brauer algebra $D_n(d)$.
$P_\ast$ is an idempotent of the $TL_n(d)$ algebra since it
satisfies \eq P_\ast P_\ast=\sum^d_{i,j,i^\prime, j^\prime} |i
i\rangle \langle j j|i^\prime i^\prime \rangle \langle j^\prime
j^\prime |=d \sum_{i,j}^d |i i\rangle \langle j j|=d P_\ast, \en
 and the permutation $P$ is a natural virtual crossing. The axiom
 $(ev/ve)$ of the Brauer algebra is satisfied since
 $P P_\ast=P_\ast P=P_\ast$, the axiom $(vee)$ can be checked via
 calculation
 \eq
 (P\otimes Id) E_2 E_1|ijk\rangle
 = (P\otimes Id)\sum_{l=1}^d |k l l \rangle \delta_{ij}
 =(Id\otimes P) E_1|ijk\rangle, \en and similarly for the verification of
 the axiom $(eev)$.

A braid representation in terms of $P_\ast$ is given by  \eq
\label{vb}
  \rho_\pm=1\!\! 1 + b_\pm P_\ast, \qquad
  b_\pm=-\frac 1 2 (d\mp \sqrt{d^2-4}), \qquad d\ge 2.
  \en
which is consistent with the case of $a=1, c=0, \lambda=d$ in
(\ref{ab}). Hence $P$ and $P_\ast$ satisfy the $(vTL)$ relation and
they generate an algebra which is the virtual TL algebra $vTL_n(d)$.

At $d=2, a=b=-1$, furthermore, $P$ and $P_\ast$ form the
unrestricted virtual TL algebra $uTL_n(2)$. The operator $1\!\!
1-P_\ast$ denoted by $P^\ast$ has the action on $|ij\rangle$ given
by
 \eq  P^\ast=1\!\! 1-\,
 P_{\ast},  \qquad
 P^{\ast} |ij\rangle =|ij\rangle- (|00\rangle+|11\rangle)
 \delta_{ij}, \qquad i,j=0,1, \en
and in terms of $P$ and $P^\ast$ it is easy to prove the following
equations,
 \eq
 \label{fu22}
  P_i P^\ast_{i+1} P^\ast_{i} = P^\ast_{i+1} P^\ast_{i} P_{i+1}, \qquad
  P^\ast_{i} P^\ast_{i+1} P_i = P_{i+1} P^\ast_{i} P^\ast_{i+1},
 \en
which are the forbidden moves $<F1>$ and $<F2>$ respectively.

Moreover,  a diagrammatical representation for permutation and its
partial transpose has been presented \cite{yong}.

\subsection{Concluding remarks and outlooks}

 In this paper, we introduce the virtual TL algebra as an extension of
 the TL algebra in view of the algebraic ansatz that the virtual TL
 algebra can represent the virtual braid group. We show that the Brauer
 algebra is the virtual TL algebra and its quotients under specific
 conditions, and present an example given by permutation and its partial
 transpose. In our further research, we will discuss virtual generalizations
 of knot invariants in terms of the trace functional over the virtual
 TL algebra and its quotients.

 We will also study the applications of the virtual TL
 algebra and virtual braid group to quantum information phenomena. In
 \cite{yong, yong1}, the $TL_n$ algebra under local unitary
 transformations is found to be a suitable algebraic structure underlying
 quantum information protocols involving maximally entangled states, and
 the teleportation configuration is recognized as a fundamental element
 in the diagrammatical representation for defining the
 the virtual TL algebra. The virtual mixed relation
 for defining the virtual braid group is a kind of formulation of
 the teleportation equation \cite{yong1}, while the virtual braid group is
 proposed to be a natural language for quantum computing \cite{yong}.

\section*{Acknowledgements}

 Y. Zhang is in part supported by NSFC grants and SRF for ROCS, SEM.

For L.H. Kauffman, most of this effort was sponsored by the Defense
Advanced Research Projects Agency (DARPA) and Air Force Research
Laboratory, Air Force Materiel Command, USAF, under agreement
F30602-01-2-05022. The U.S. Government is authorized to reproduce
and distribute reprints for Government purposes notwithstanding any
copyright annotations thereon. The views and conclusions contained
herein are those of the authors and should not be interpreted as
necessarily representing the official policies or endorsements,
either expressed or implied, of the Defense Advanced Research
Projects Agency, the Air Force Research Laboratory, or the U.S.
Government. (Copyright 2006.) It gives L.H. Kauffman great pleasure
to acknowledge support from NSF Grant DMS-0245588.

 \end{document}